\documentclass[12pt]{article}
\usepackage{amsfonts,amssymb,amsmath}

\title{Response to Horton and Dewdney}
\author{
	Roderich Tumulka\footnote{address: Mathematisches Institut 
	der Universit\"at M\"unchen, Theresienstr.\ 39, D-80333 M\"unchen, 
	Germany. E-mail: tumulka@mathematik.uni-muenchen.de}
	}
\date{}

\begin{document}
\maketitle

\begin{abstract}
  There are some points in the reply of Horton \textit{et al.}\ 
  \cite{reply} to my comment \cite{comment} on their paper
  \cite{Dewdney2} which I cannot let stand without a response. I
  provide here some clarification of how much I proved about the set
  of points where their law of motion is ill-defined.
\end{abstract}
\openup-.1\jot

In a recent article in J.\ Phys.\ A \cite{Dewdney2}, Horton \textit{et
al.}\ present what they claim is a Bohm-type law of motion for point
particles, based on a Klein--Gordon wave function and implying (unlike
a similar law proposed by de~Broglie) timelike world lines. Concerning
this claim I pointed out in a comment \cite{comment} that the
prescription they give is ill-defined in some situations, and
underpinned this by a concrete example. In addition, I gave arguments
to the effect that the set of ``bad'' space-time points, where the law
of motion is ill-defined, is a set of positive measure for many wave
functions. To this Horton \textit{et al.}\ have responded
\cite{reply}, ignoring my arguments, that although bad points may
exist, they form a set of lesser dimension and therefore can be dealt
with by a limiting procedure. I wish here to point out that the
response of Horton \textit{et al.}\ is entirely without merit. Here is
why:

In my comment I pointed out that those space-time points are bad where
both vectors $W^+_\mu$ and $W^-_\mu$ that appear in the law of Horton
\textit{et al.}\ are spacelike, or, equivalently, where $W^+_\mu
W^{+\mu} < 0$ and $W^-_\mu W^{-\mu} < 0$. Given that the vector fields
$P_\mu$ and $S_\mu$ (on which the construction of $W^+_\mu$ and
$W^-_\mu$ relies) are continuous, the functions $W^+_\mu W^{+\mu}$ and
$W^-_\mu W^{-\mu}$ are continuous, too, and thus their values remain
negative in an \emph{entire neighborhood} of any bad space-time point.

Therefore, the bad points form an open set, quite contrary to the
picture of ``nodal lines'' or even ``isolated points'' that Horton
\textit{et al.}\ suggest in their reply \cite{reply}. This is also
true of my specific example wave function of positive energy (having
continuous $P_\mu$ and $S_\mu$), in spite of their claim to the
contrary. Thus, my example is as well a counter example to the claim
that the bad points form a set of lesser dimension. As a consequence,
the limiting process they suggest cannot be carried out.

Horton \textit{et al.}\ are mistaken when they write, ``Tumulka,
however, goes on to conjecture, but not prove that ... the set of
pairs $P_\mu, S_\mu$ where both $W^+_\mu$ and $W^-_\mu$ are spacelike
is open in the 8-dimensional space of the pairs'' $P_\mu, S_\mu$. I
did give a proof for that---in the very sentence of my comment that
Horton \textit{et al.}\ quote.

Horton \textit{et al.}\ write, ``Tumulka provides neither proof nor
grounds for his expectation'' that ``the space-time points with
spacelike $W^+_\mu$ and $W^-_\mu$ form a set of positive measure.''
Since every nonempty open set has positive measure, my claim follows
directly from the very reasonable expectation that $P_\mu$ and $S_\mu$
are continuous.

Horton \textit{et al.}\ write, ``A straightforward computation shows
that the points where $S_0 =0$ form nodal lines as do the points where
$P_0=0$." I emphasize that the relevant points are not merely those
where $S_0=0$ and $P_0=0$, but many more, namely all those where
$S_\mu$ and $P_\mu$ span a spacelike 2-plane. The latter condition is
indeed equivalent to $S'_0=0$ and $P'_0=0$ with respect to \emph{some}
Lorentz frame, but this is a condition very different from $S_0=0$ and
$P_0=0$ in one \emph{fixed} frame. Conflating these two, Horton
\textit{et al.}\ present a computation that fails to take into
account that $v_k$, the parameter of the Lorentz boost that will make
$S'_0(x^\mu)$ and $P'_0(x^\mu)$ zero, may well depend on $x^\mu$.
Thus, the computation they do is completely irrelevant.

\end{document}